\documentclass[aip,apl,preprint]{revtex4-1}

\usepackage{graphicx}
\usepackage{dcolumn}
\usepackage{bm}
\usepackage{xspace, epstopdf}
\usepackage{amsmath}    

\begin{document}

\title{Gate voltage control over spin relaxation length}
\author{S. M. Frolov}
\affiliation{Department of Physics and Astronomy, University of British Columbia, Vancouver, British Columbia V6T 1Z4, Canada}
\affiliation{Kavli Institute of Nanoscience, Delft University of Technology, 2600 GA Delft, The Netherlands}
\author{W.W. Yu}
\affiliation{Department of Physics and Astronomy, University of British Columbia, Vancouver, British Columbia V6T 1Z4, Canada}
\author{S. L\"uscher}
\affiliation{Department of Physics and Astronomy, University of British Columbia, Vancouver, British Columbia V6T 1Z4, Canada}
\author{J. A. Folk}
\affiliation{Department of Physics and Astronomy, University of British Columbia, Vancouver, British Columbia V6T 1Z4, Canada}
\author{W. Wegscheider}
\affiliation{Laboratorium f\"ur Festk\"orperphysik, ETH Z\"urich, 8093 Z\"urich, Switzerland}
\date{\today}

\begin{abstract}

Spin currents in channels of a high mobility GaAs/AlGaAs two-dimensional electron gas are generated and detected using spin-polarized quantum point contacts. We have recently shown that the relaxation length of spin currents is resonantly suppressed when the frequency at which electrons bounce between channel walls matches the Larmor frequency. Here we demonstrate that a gate on top of the channel tunes such ballistic spin resonance by tuning the velocity of electrons and hence the bouncing frequency. These findings demonstrate a new mechanism for  electrical control of spin logic circuits.

\end{abstract}

\maketitle

In a spin field-effect transistor, a spin current is controlled by voltages on a gate electrode \cite{ZuticRMP04, SugaharaIEEE2010}. The primary means for coupling spin to electric fields in a semiconductor is spin-orbit interaction \cite{Winkler}, through which electron spins experience an effective magnetic field that is momentum dependent.  As electrons travel through a semiconductor crystal, their spins precess under the influence of this intrinsic spin-orbit field. One route to making a spin transistor is to use the electric field to tune the strength of spin-orbit interaction \cite{DattaAPL90, SchliemannPRL03, CartoixaAPL03, KooScience09}. However, this strategy requires either a substantial change in the electric field or density, or a strong spin-orbit anisotropy.

Here we demonstrate a different paradigm for making a spin transistor, in which spin current relaxation length is resonantly suppressed using only a small change in gate voltage. The gate controls the Fermi velocity of electrons, and the device dimensions, while the intrinsic spin-orbit interaction strength is fixed. Spin-polarized electrons experience a periodic spin-orbit field as they bounce through channels of a high mobility two-dimensional electron gas (2DEG). Spin relaxation is enhanced when the frequency of trajectories bouncing between the channel walls matches the spin precession frequency set by external magnetic field \cite{FrolovNature09}. The relaxation length can be suppressed by an order of magnitude using this method,\cite{Yuarxiv10} and spin accumulation is exponential in spin relaxation length. As the bouncing frequency changes with the gate voltage, spin dynamics can be tuned on and off of resonance.  Because this effect is based on a Fermi velocity resonance rather than a direct modification of the barrier to electron flow, as occurs in a conventional transistor, small changes in gate voltage can in principle yield very large changes in the spin signal.

Our devices are created in a GaAs/AlGaAs 2DEG (electron density $n_s=1.11\times 10^{11}$ cm$^{-2}$ and mobility $\mu=4.44\times 10^{6}$ cm$^2$/Vs measured at $T=1.5$ K) using electrostatic surface gates (Figs. 1(a), 1(b)). Each device consists of a spin injector and a detector located along a spin diffusion channel of width w = 2.5 $\mu$m \cite{FrolovPRL09}. The spin-polarized charge current from the injector is routed towards the left end of the channel, which is electrically grounded. A pure spin current is driven by diffusion into the right part of the channel, which is electrically floating \cite{JohnsonPRL85, JedemaNature01}.  A 500 nm-wide top-gate runs through the middle of the channel along its length.  It is this gate that changes the relaxation length of spin currents, by tuning the velocity of electrons as they traverse the middle section of the channel. An additional gate at the right end of the channel ($\Lambda$-gate) is used to measure spin relaxation length \cite{FrolovPRL09}. 

\begin{figure*} 
 \includegraphics[scale=1]{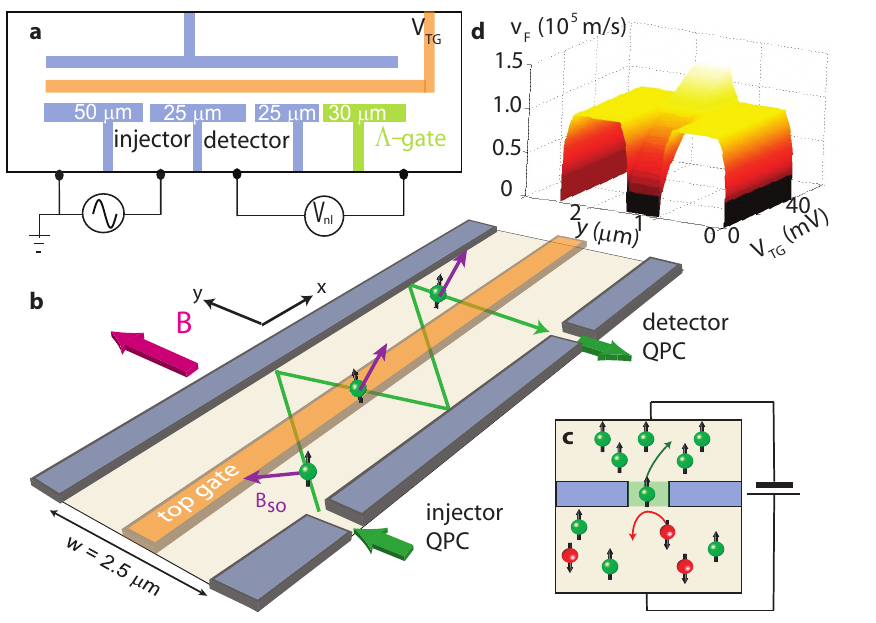}

\caption{ (a) device schematic showing gates that define injector and detector QPCs and the spin current propagation channel (blue), $\Lambda$-gate (green) and top gate (orange). Gates are not to scale, their lengths are indicated. (b) spin-polarized electrons injected into the channel propagate along ballistic trajectories (green) under the influence of fields $B$ and $B_{so}$. (c) illustration of spin filter in a QPC: spin-up electrons (green) are transmitted through the QPC, while spin-down electrons (red) are reflected. (d) Fermi velocity modeled along y-axis for a range of top gate voltage. }\end{figure*}

The injector/detector part of the device contains two quantum point contacts (QPCs): 200 nm gaps in the channel walls, which can be spin selective in high magnetic fields. QPCs transmit an integer number of quasi-one-dimensional modes \cite{vanWeesPRL88}. At low field their gate-tuneable conductance is quantized in units of $2e^2/h$, where factor 2 comes from spin degeneracy. At high magnetic fields, $B>$3 T, spin degeneracy is lifted and additional conductance plateaus develop at odd multiples of $e^2/h$. At the $1e^2/h$ plateau, QPCs transmit only spin-up electrons, while spin-down electrons are reflected (Fig. 1(c)).

Spin currents are generated by passing a charge current through a QPC; in this case the injected spins are polarized along the external magnetic field. Another QPC located 25 $\mu$m to the right is used as a detector of pure spin currents. The measure of spin current is the nonlocal voltage $V_{nl}$ which develops across the detector QPC: the pure spin current that diffuses down the channel from the injector drives spin-up electrons at this location to a chemical potential above equilibrium, and spin-down electrons to a chemical potential below equilibrium.  For the experiments described here, both QPCs are set to $1e^2/h$ plateaus for maximum polarization and hence the maximum injection and detection efficiency.

The effect that underlies gate-tuneable spin relaxation is termed ballistic spin resonance (BSR) \cite{FrolovNature09}. The mean free path in the spin diffusion channels is around 10 $\mu$m,  much longer than the channel width. This means that trajectories of injected electrons involve multiple specular reflections from channel walls: periodic trajectories of the type illustrated in Fig. 1(b).  BSR is driven by this ballistic motion, which is converted by spin-orbit interaction into a periodic magnetic field.  For a channel aligned along the x-axis, as in Fig. 1(b), the x-component of the electron momentum is preserved through specular reflections while the y-component changes sign periodically in time: a square-wave pattern with frequency components at odd harmonics of the primary bouncing frequency $v_F/2w$, where $v_F$ is the Fermi velocity.

An oscillating y-component of momentum gives rise to an oscillating x-component of the spin-orbit field, $B_{so}^x$, due to the symmetry of the spin-orbit interaction in these materials.\cite{FrolovNature09}  The external magnetic field $B$, applied perpendicular to the channel (that is, applied along the y-axis), sets the spin precession frequency $g\mu_B B/ h$ and drives spin resonance when $g\mu_B B/ h \approx v_F/2w$. On resonance the y-projection of spin changes sign after several bounces. Since every electron follows a different bouncing trajectory the injected spin polarization is rapidly randomized and the spin current propagation length is decreased. As a consequence, BSR manifests itself in a dip in the spin signal $V_{nl}$.

The density of  electrons, and hence $v_F$, can be changed in the central region of the channel using a voltage on the top gate, V$_{TG}$; this effect is used to tune in and out of ballistic spin resonance. As illustrated in Fig. 1(d), the velocity underneath the top gate can be varied from zero to a level exceeding that of the ungated 2DEG.  A change in the velocity of electrons translates into a change in the BSR frequency, which shifts the BSR dip position in magnetic field (Fig. 2).  Figure 2(a) shows the evolution of the first harmonic BSR(1) and the third harmonic BSR(3) with top gate voltage.  For more positive top-gate voltages the bouncing frequency increases, which leads to a BSR shift towards higher field. For V$_{TG}<$-150 mV, on the other hand the 2DEG is entirely depleted underneath the top gate.

\begin{figure*} 
 \includegraphics[scale= 1]{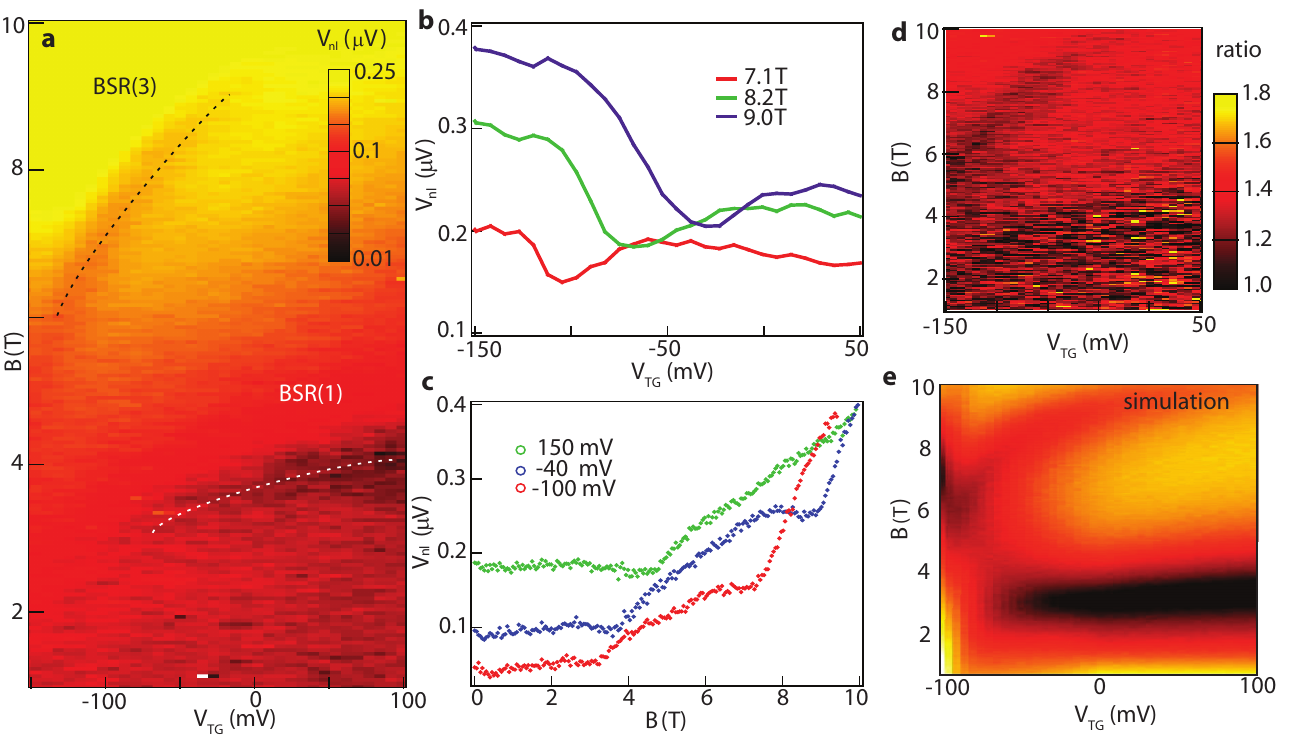}

\caption{ (a) Device fabricated along [-110]. Nonlocal signal as a function of top gate and  magnetic field. BSR(1) and BSR(3) mark first and third harmonic BSR positions. (b) and (c) show cuts from (a). (d) Ratio of nonlocal signals for $\Lambda$-gate depleted and undepleted characterizes changes in spin current relaxation length. (e) Monte Carlo simulations of non-local spin signal assuming full QPC polarization (in arbitrary units, yellow indicates large spin signal).}\end{figure*}

From a spintronics point of view it is practical to fix the external magnetic field and tune spin current with V$_{TG}$ (Fig. 2(b)). As the voltage is tuned, changes in the spin signal $V_{nl}$ reflect a change in spin relaxation length at the dip corresponding to BSR(3). (The resistivity of the channel changes by less than 5$\%$ over this gate voltage range.) Fig. 2(c) shows the magnetic field evolution of the spin signal. The overall increase of spin signal at higher magnetic fields is due to the rising polarization of the injector and detector QPCs, which is relatively low in these QPCs below 6 T. At low magnetic fields $B<4$ T the QPC polarization is very small and the signal contains a significant contribution from thermal effects.\cite{FrolovPRL09} These factors reduce the contrast of BSR(1), which for these wide channels appears at low magnetic fields.

In order to confirm that the observed feature is due to a change in spin relaxation length and not, for example, a change in the injector or detector polarization, we vary the channel length in-situ by depleting and undepleting the $\Lambda$-gate \cite{FrolovPRL09}. The ratio of spin signal for the $\Lambda$-gate depleted/undepleted is suppressed when V$_{TG}$ and B are set to BSR(1) or BSR(3) (see Figs. 2(d) and 2(a)). This is as expected from a simple diffusion model: a shorter spin current propagation length means that manipulations with a far-away $\Lambda$-gate have less influence on the nonlocal signal.  The data are closely reproduced by Monte Carlo simulations of spin dynamics in a top gated channel, where the local electron momentum depends on the electrostatic environment induced by the gates (Fig. 2(e)) \cite{LuscherPRB10}. The simulations use the geometry of Fig. 1(b), the density profile of Fig. 1(d) and 2DEG parameters determined in previous publications\cite{FrolovPRL09, FrolovNature09, Yuarxiv10}. 

Data in Fig. 2 were obtained on a device fabricated along the [-110] crystal axis. The oscillating spin-orbit field is then due to bouncing along the [110] axis, where the spin-orbit interaction is substantially weaker due to a cancellation of terms originating from bulk-inversion asymmetry and structure-inversion asymmetry \cite{Yuarxiv10}. As a result of the weak spin-orbit driving field, the on/off ratio for spin current is only a factor of two in this device. At the same time, this weak intrinsic driving results in narrower BSR dips, equivalent to the narrowing of features for weaker transverse fields in nuclear or electron spin resonance. 

\begin{figure*} 
 \includegraphics[width= 0.7\textwidth]{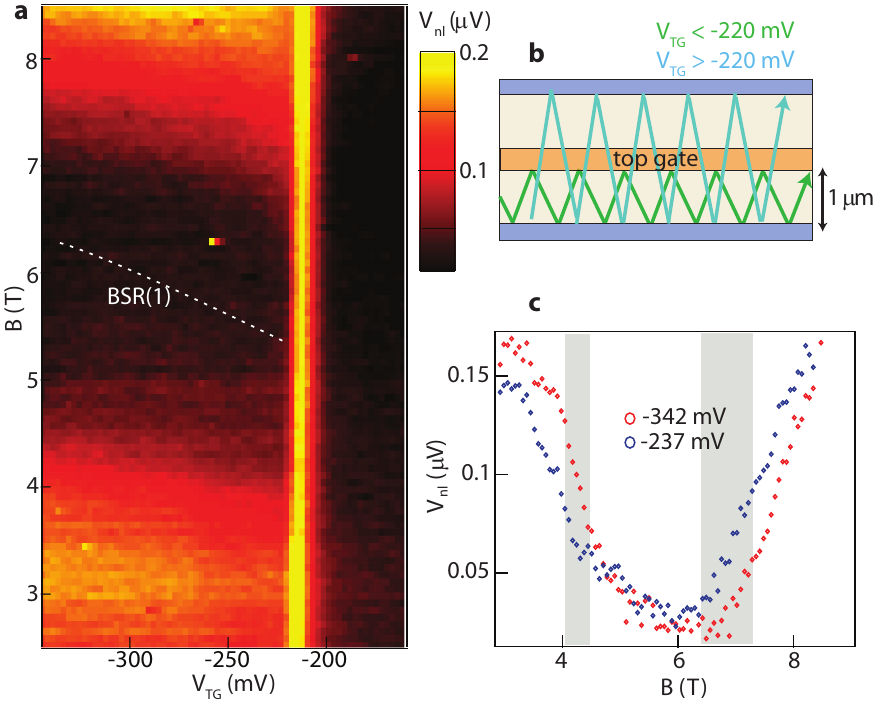}

\caption{ (a) Device fabricated along [110]. Nonlocal signal as a function of top gate and magnetic field. At $V_{TG}$ = -220 mV the 2DEG underneath the top gate becomes depleted. BSR(1) marks the first harmonic BSR position. (b) schematic of ballistic trajectories confined to the lower part of the channel due to depleted top gate (green) and trajectories spanning the whole channel width when the top gate is undepleted (light blue). Top gate voltage ranges corresponding to both trajectories are given above. (c) linecuts from (a), gray areas highlight regions with high on/off ratio. }\end{figure*}

Dramatically higher on/off ratios can be obtained in channels fabricated along the orthogonal [110] axis. In this case, electrons bounce along [-110], where the two spin-orbit fields add and the total spin-orbit field is greatly enhanced \cite{Yuarxiv10}. Fig. 3(a) shows data analogous to that shown in Fig. 2(a) but for a device fabricated along [110]. With the top gate undepleted (V$_{TG}>$  -220 mV, 2.5 micron wide channel), as was the case for  Fig.~2, spin accumulation is negligible for any values of B because the oscillating spin-orbit field is strong, giving BSR(1) and BSR(3) dips that are much broader in field and overlap to cover the entire range of magnetic field \cite{FrolovNature09}. When the top gate is fully depleted, on the other hand, the effective channel width is 1 micron and a pronounced dip at BSR(1) is observed at B$\approx$6 T. The strong suppression of the spin signal in the center of the dip indicates that the spin relaxation length is much shorter than the injector-detector spacing of 25 microns.

With electrons confined to only the lower part of the channel, we demonstrate another mode of field-effect control over spin current (Fig. 3(b)). For a more negative top gate the effective channel width is decreased, which in turn increases the bouncing frequency and pushes the BSR(1) dip  to higher magnetic fields.  Figure 3(c) shows that significant on-off ratios of the spin signal can be achieved within narrow regions of magnetic field.  Improved on/off ratios over a much broader field range could be achieved by covering the entire channel with a top gate. In this case the position of the BSR dip would be much more sensitive to the top gate voltage because the velocity is tuned over the entire trajectory.

In conclusion, gate-tuneability of ballistic spin resonance provides a means of field effect control over spin current propagation in two-dimensional electron gases. While at present spin current can only be suppressed, in the future coherent spin rotations can be controlled using a top gate if all injected electrons follow the same trajectory, for instance in a transverse electron focusing geometry.

\bibliographystyle{unsrt}
\bibliography{BSR}

\begin{thebibliography}{10}

\bibitem{ZuticRMP04}
I.~\ifmmode \check{Z}\else \v{Z}\fi{}uti\ifmmode~\acute{c}\else \'{c}\fi{},
  J.~Fabian, and S.~Das~Sarma.
\newblock Spintronics: Fundamentals and applications.
\newblock {\em Rev. Mod. Phys.}, 76(2):323--410, 2004.

\bibitem{SugaharaIEEE2010}
S.~Sugahara and J.~Nitta.
\newblock Spin-transistor electronis: Overview and outlook.
\newblock {\em Proceedings of the IEEE}, 98(12):2124, 2010.

\bibitem{Winkler}
R.~Winkler.
\newblock {\em Spin-Orbit Coupling Effects in Two-Dimensional Electron and Hole
  Systems}.
\newblock Springer, Berlin, 2003.

\bibitem{DattaAPL90}
S.~Datta and B.~Das.
\newblock Electronic analog of the electro‐optic modulator.
\newblock {\em Appl. Phys. Lett.}, 56:665, 1990.

\bibitem{SchliemannPRL03}
J.~Schliemann, J.~C. Egues, and D.~Loss.
\newblock Nonballistic spin-field-effect transistor.
\newblock {\em Phys. Rev. Lett.}, 90:146801, Apr 2003.

\bibitem{CartoixaAPL03}
X.~Cartoixa, D.~Z.-Y. Ting, and Y.-C. Chang.
\newblock A resonant spin lifetime transistor.
\newblock {\em Applied Physics Letters}, 83(7):1462--1464, 2003.

\bibitem{KooScience09}
H.~C. Koo, J.~H. Kwon, J.~Eom, J.~Chang, S.~H. Han, and M.~Johnson.
\newblock Control of spin precession in a spin-injected field effect
  transistor.
\newblock {\em Science}, 325(5947):1515--1518, 2009.

\bibitem{FrolovNature09}
S.~M. Frolov, S.~L\"uscher, W.~Yu, Y.~Ren, J.~A. Folk, and W.~Wegscheider.
\newblock Ballistic spin resonance.
\newblock {\em Nature}, 458(7240):868--871, 2009.

\bibitem{Yuarxiv10}
W.~W. Yu, S.~M. Frolov, S.~L\"uscher, J.~A. Folk, and W.~Wegscheider.
\newblock Spin-orbit anisotropy measured using ballistic spin resonance.
\newblock {\em arXiv}, page 1009.5702, 2010.

\bibitem{FrolovPRL09}
S.~M. Frolov, A.~Venkatesan, W.~Yu, J.~A. Folk, and W~Wegscheider.
\newblock Electrical generation of pure spin currents in a two-dimensional
  electron gas.
\newblock {\em Phys. Rev. Lett.}, 102(11):116802, 2009.

\bibitem{JohnsonPRL85}
M.~Johnson and R.~H. Silsbee.
\newblock Interfacial charge-spin coupling - injection and detection of spin
  magnetization in metals.
\newblock {\em Physical Review Letters}, 55(17):1790--1793, 1985.

\bibitem{JedemaNature01}
F.~J. Jedema, A.~T. Filip, and B.~J. van Wees.
\newblock Electrical spin injection and accumulation at room temperature in an
  all-metal mesoscopic spin valve.
\newblock {\em Nature}, 410(6826):345--348, 2001.

\bibitem{vanWeesPRL88}
B.~J. van Wees, H.~van Houten, C.~W.~J. Beenakker, J.~G. Williamson, L.~P.
  Kouwenhoven, D.~van~der Marel, and C.~T. Foxon.
\newblock Quantized conductance of point contacts in a two-dimensional
  electron-gas.
\newblock {\em Physical Review Letters}, 60(9):848--850, 1988.

\bibitem{LuscherPRB10}
S.~L\"uscher, S.~M. Frolov, and J.~A. Folk.
\newblock Numerical study of resonant spin relaxation in quasi-one-dimensional
  channels.
\newblock {\em Phys. Rev. B}, 82:115304, Sep 2010.

\end{thebibliography}

 \end{document}